# Prediction of half-metallic properties in TlCrS$_2$ and TlCrSe$_2$ based on density functional theory

F.M.Hashimzade[1], D.A.Huseinova[1], Z.A.Jahangirli[1,2], B.H.Mehdiyev[1]

[1] Institute of Physics, National Academy of Sciences of Azerbaijan, AZ 1143, Baku, Azerbaijan

[2] Institute of Radiaton Problems, National Academy of Sciences of Azerbaijan, AZ 1143, Baku, Azerbaijan

**ABSTRACT**

Half-metallic properties of TlCrS$_2$, TlCrSe$_2$ and hypothetical TlCrSSe have been investigated by first-principles all-electron full-potential linearized augmented plane wave plus local orbital (FP-LAPW+lo) method based on density functional theory (DFT). The results of calculations show that TlCrS$_2$ and TlCrSSe are half-metals with energy gap (E$_g$) ~0.12 ev for spin-down channel. Strong hybridization of p-state of chalchogen and d-state of Cr leads to bonding and antibonding states and subsequently to the appearance of a gap in spin-down channel of TlCrS$_2$ and TlCrSSe. In the case of TlCrSe$_2$, there is a partial hybridization and p-state is partially present in the DOS at Fermi level making this compound nearly half- metallic. The present calculations revealed that total magnetic moment keeps its integer value on a relatively wide range of changes in volume (-10% ÷ 10%) for TlCrS$_2$ and TlCrSSe, while total magnetic moment of TlCrSe$_2$ decreases with increasing volume approaching to integer value 3μB.

## 1. Introduction

Searching for new half-metallic materials with 100% spin polarization is an actual problem of material science in connection with the future development of spintronics as an alternative to the electronics. Because of the experimental difficulties to establish half-metallicity, the main method searching for new half-metallic ferromagnetic materials nowadays are electronic structure calculations [1]. We assume that the compound TlCrS$_2$ and TlCrSe$_2$, which according to [2,3] are ferromagnetic, may exhibit the half-metallic ferromagnetic properties. The integer total magnetic moments can be accounted using Slater-Pauling rule modified by Kübler [4]. In the semiconductor compounds $A^{III} B^{III} C_2^{VI}$ eighteen electrons completely fill nine valence bands. Thus, the number of occupied bands in semiconducting channel of TlCrS$_2$ is nine. The total number of electron/unit-cell is twenty one and from the Slater-Pauling rule the total magnetic moment of TlCrS$_2$ should be 3 μB/unit-cell.

Results of our calculations have indeed shown that TlCrS$_2$ to be half-metallic ferromagnet. The same holds for the hypothetical TlCrSSe compounds which is obtained from TlCrS$_2$ by replacing S atoms with the Se atoms in one of the sublattices.

## 2. Structural properties

TlCrS$_2$ have rhombohedral symmetry with space group R3m, space group number 160. The unit cell of the crystal contains one formula unit. The structure of TlCrS$_2$ is shown in Fig. 1 (From [2], Fig.1). The structure consists of atomic layer arranged in sequence: Cr-S-Tl-S. Thallium and chromium atoms are in trigonal prismatic coordinated, surrounded by sulphur atoms. In the hexagonal setting lattice parameters are a= 3.538 Å and c= 21.92 Å. Wyckoff positions of all atoms are 3a: (0,0,z), (2/3,1/3,1/3+z),(1/3,2/3,2/3+z) with z(Tl)=1/6, z(Cr)=0, z(S)=±0.39. For the isostructural TlCrSe$_2$ compound lattice parameters are a = 3.700Å, c = 22.68Å [2]. There is no any information present in literature about Wyckoff positions of atoms for TlCrSe$_2$.

## 3. Computational details

The geometry optimization and electronic structure calculations were performed in the spin polarized DFT [5,6] framework based on the FP-LAPW +lo method as implemented in the Wien2k program package [7]. The Kohn–Sham single particle equations are solved for both spin-up and spin-down densities self-consistently. The exchange-correlation potential was calculated within the general gradient approximation (GGA) using scheme suggested by Perdew et al. [8]. The influence of relativistic effects on the structural and electronic properties of the crystals are taken into account in the scalar approximation neglecting spin–orbit coupling. The process of minimization continued until the force moduli became less than ~0.001 Ry/Bohr. In our calculations, the convergence parameter $R_{mt}K_{max}$ was set to 7.0, where $R_{mt}$ and $K_{max}$ are the smallest of the muffin–tin sphere radii and the largest reciprocal lattice vector used in the plane wave expansion, respectively. Inside atomic spheres the partial waves were expanded up to $l_{max}$ = 10. Integrations in reciprocal space were performed using the tetrahedron [9] method with 1000 k points in the first Brillouin zone to determine the charge density in each self-consistency step. The muffin-tin sphere radii $R_{mt}$ =2.5 Bohr, 2.4 Bohr, 2.06 Bohr and 2.37 Bohr were used for Tl, Cr, S and Se, respectively. The cut-off energy for separation of the core and valence states was set to -6.0 Ry. Self-consistency is achieved with energy convergence of $10^{-5}$ eV. Due to the localized feature of 3d-electrons of Cr, the on-site Coulomb interaction was taken into account for the correct description of this material [10]. In our calculations we used U=2.7 eV.

## 4. Results and discussions

To study the electronic structure and magnetic properties of TlCrS$_2$ and TlCrSe$_2$ we first fully relaxed atomic positions until forces on each atom were below ~0.001 Ry/Bohr and lattice constants by minimizing the total energy for various lattice parameters. Minimization was performed imposing additional condition c/a=const. Since there is no information in literature about the internal positional parameters of TlCrSe$_2$, they have been obtained by relaxing the internal parameters of TlCrS$_2$. The values of the optimized lattice parameters are given in Table. 1.

*Table1. Predicted equilibrium lattice parameter and Wyckoff positions*

| compounds | a (Å) | c (Å) | z(Tl) | z(Cr) | z(S(Se))1 | z(S(Se))2 |
|---|---|---|---|---|---|---|
| TlCrS2 | 3.522 | 21.822 | 0.1651 | 0.0000 | 0.3900 | 0.6107 |
| TlCrSe2 | 3.6621 | 22.4475 | 0.1661 | 0.0000 | 0.3949 | 0.6057 |
| TlCrSSe | 3.6623 | 22.4448 | 0.1547 | 0.0000 | 0.3813 | 0.6036 |

Calculated total energy of TlCrS$_2$, TlCrSe$_2$ and TlCrSSe compounds as function of the unit cell volume shown in Fig. 2-4. The minimum of the total energy corresponds to the equilibrium value of the unit cell volume at zero temperature and pressure. A energy-vs-volume function E(V) has been fitted using the Murnaghan equation of state. The minimum of this function provides the equilibrium volume and bulk modulus for the TlCrS$_2$, TlCrSe$_2$ and TlCrSSe compounds [11]:

$$E_{tot}(V) = E_{tot}(V_0) + (BV/B_P(B_P - 1))\left(B_P(1 - V_0/V) + \left(\frac{V_0}{V}\right)^{B_P} - 1\right) \quad (1)$$

where $V_0$ is the equilibrium volume, $E_{tot}(V_0)$ is the equilibrium energy, B the bulk modulus, and $B_P$ is the first derivative of B with respect to pressure.

Fig. 5-7 shows the spin-polarized total and orbital-decomposed density of states of TlCrS$_2$, TlCrSe$_2$ and TlCrSSe for the spin-down and the spin-up channels at optimized lattice constants. As the crystal structures are identical there are similarity between the DOS of these compounds. For all compounds the spin-up channel is metallic which has non-zero DOS at the Fermi level ($E_F$) whereas in the spin-down channel there is small energy gap ~ 0.12 eV for TlCrS$_2$ and TlCrSSe. Thus these compounds are truly half-metallic ferromagnetic. For TlCrSe$_2$ there is non-zero DOS at $E_F$ making this compound nearly half- metallic. The value of electron-spin polarization at Fermi level is of fundamental importance for the use of a compound in scientific and technological applications. Spin polarization P for ferromagnetic material is given by

$$P = (N\uparrow(E_F) - N\downarrow(E_F))/(N\uparrow(E_F) + N\downarrow(E_F)) \quad (2)$$

where $N\uparrow(E_F)$ and $N\downarrow(E_F)$ are the spin dependent density of states at the Fermi level. The contribution from p-state S(Se) to spin-up DOS for TlCrS$_2$, TlCrSe$_2$ and TlCrSSe two times larger than from the d-state of Cr. The contribution from d-state of Cr to spin-down DOS practically absent. Very small spin-down DOS at Fermi level for TlCrSe2 is due to the presence of p-state of Se. The presence of p-states of the Se atom at the Fermi level in the case of spin-down channel gives rise to the pseudogap and therefore leads to no 100% spin polarization.

As can be seen from the spin-down DOS of TlCrS$_2$ p-state of S and d-state of Cr equally extended over a large energy range, which increases the hybridization. Strong hybridization of these states leads to bonding and antibonding states and therefore to the appearance of a gap in spin-down channel of TlCrS$_2$ and TlCrSSe. In the case of TlCrSe$_2$, there is a partial hybridization and p-state is partially present in the DOS at E$_F$.

The calculated total, atom-resolved and interstitial magnetic moments, values of spin polarization for TlCrS$_2$, TlCrSe$_2$ and TlCrSSe have been shown in Table 2. As can be seen from table the largest contribution to the total magnetic moment comes from Cr atoms. This is due to the large exchange splitting between spin-up and spin-down states of Cr atoms. The contribution of Tl and chalchogen atoms to the total magnetic moment is small.

*Table2. Spin magnetic moments*

| Compounds | $M_{int}$ | $M_{Tl}$ | $M_{Cr}$ | $M_{S1}$ | $M_{S2}$ | $M_{cell}$ | P |
|---|---|---|---|---|---|---|---|
| TlCrS2 | 0.1324 | 0.0004 | 3.0124 | -0.0736 | -0.0711 | 3.0004 | 100% |
| TlCrSe2 | 0.1450 | -0.0103 | 3.1220 | -01200 | -0.1185 | 3.0181 | 98% |
| TlCrSSe | 0.1351 | -0.0002 | 3.0881 | -0.0867 | -0.1360 | 3.0001 | 100% |

The total magnetic moment obtained from spin polarized calculations is 3 µB for TlCrS$_2$ and TlCrSSe, but for TlCrSe$_2$ total magnetic moment slightly deviate from integer value. Spin polarized results predict strictly half-metallic behavior for TlCrS$_2$ and TlCrSSe.  We suppose that this compounds may be valuable for spintronic application. As is known, integer value of the total magnetic moment (in Bohr magneton) is a necessary but not sufficient condition for the existence of half-metallic state. For a 100% spin polarization it is necessary to have the energy gap in the minor spin states [12,13]. Therefore, in each case it is necessary to check the presence of a non-zero spin resolved DOS at the Fermi level. In our case, the density of states in the spin-down channel at the Fermi level for TlCrSe$_2$ is small, which leads to no 100% spin polarization. Changing lattice parameters of this compound, for example with pressure can be achieved truly half-metallicity.

The variation of the structural parameters substantially changes the electronic properties of the crystals. It is, therefore, important to investigate the dependence of the half-metallicity of TlCrS$_2$, TlCrSe$_2$ and TlCrSSe with respect to the unit cell volume. The dependence of total magnetic moment on unit cell volume are shown in Fig. 8-10 for all compounds. As can be seen from Fig. 8 and 10  total magnetic moment keeps its integer value on a relatively wide range of changes in volume  (-10% ÷ 10%) for TlCrS$_2$ and TlCrSSe, while total magnetic moment of TlCrSe$_2$ decreases with increasing volume, approaching to integer value of 3µB. Also be seen that the partial spin moments of Cr atom increases with increasing volume. It is because the hybridization between neighboring atomic states decreases with increasing lattice constant and d-states acquire more atomic-like character resulting in increase of the partial spin moments of Cr atom.

5. Conclusion

We have investigated magnetic properties of TlCrS$_2$, TlCrSe$_2$ and TlCrSSe from first-principles. Our calculations predict that TlCrS$_2$ and TlCrSSe are half-metals with E$_g$ ~0.12 eV for spin-down channel. The states at the Fermi level in the case of spin-up channel consist of a hybridization of p-states of the atom S(Se) with d-states of Cr. The presence of Se p-states at the Fermi level for TlCrSe$_2$ in the case of spin-down channel gives rise to the pseudogap and therefore leads to no 100% spin polarization. Strong hybridization of p-state of chalchogen and d-state of Cr leads to bonding and antibonding states and therefore to the appearance of a gap in spin-down channel of TlCrS$_2$ and TlCrSSe. In the case of TlCrSe$_2$, there is a partial hybridization and p-state is partially present in the DOS at E$_F$ making this compound nearly half- metallic. Total magnetic moment keeps its integer value on a relatively wide range of changes in volume (-10% ÷ 10%) for TlCrS$_2$ and TlCrSSe, while total magnetic moment TlCrSe$_2$ decreases with increasing volume, approaching to integer value 3μB. We believe that these compounds are promising materials in spintronics.

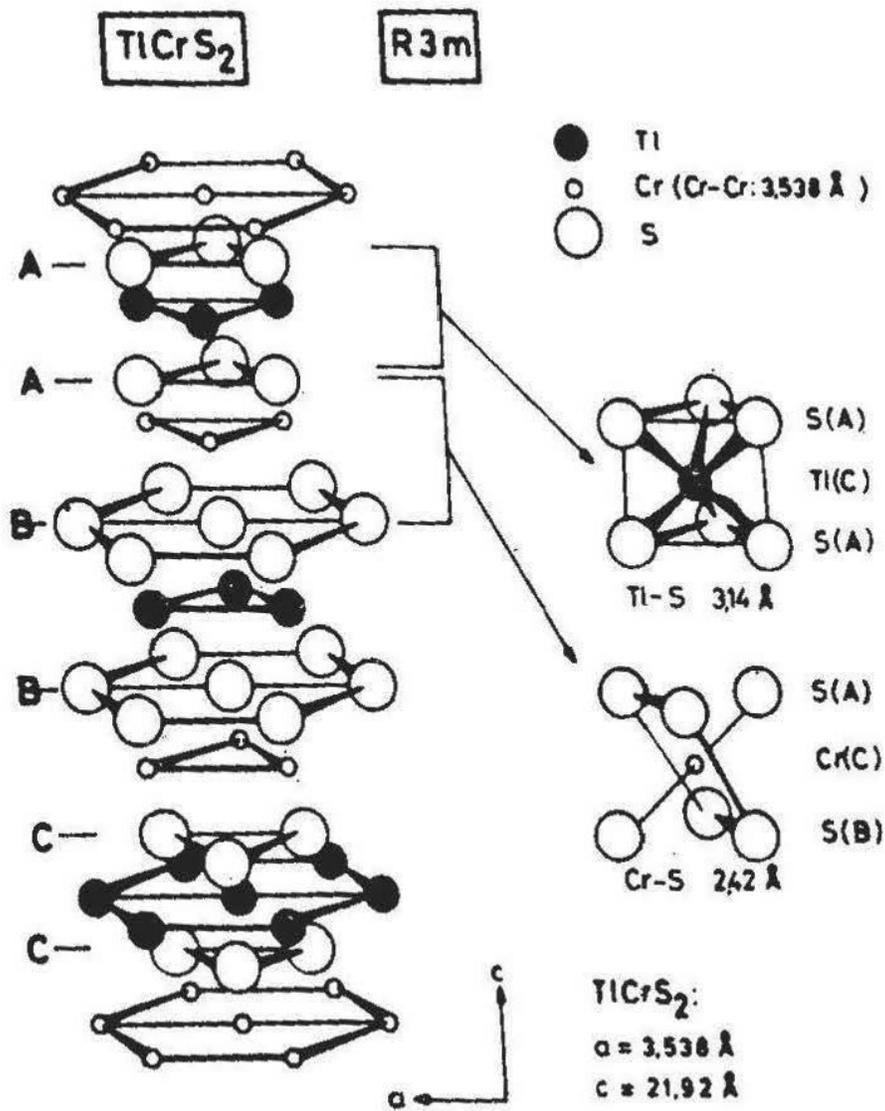

Fig.1. TlCrS$_2$ structure

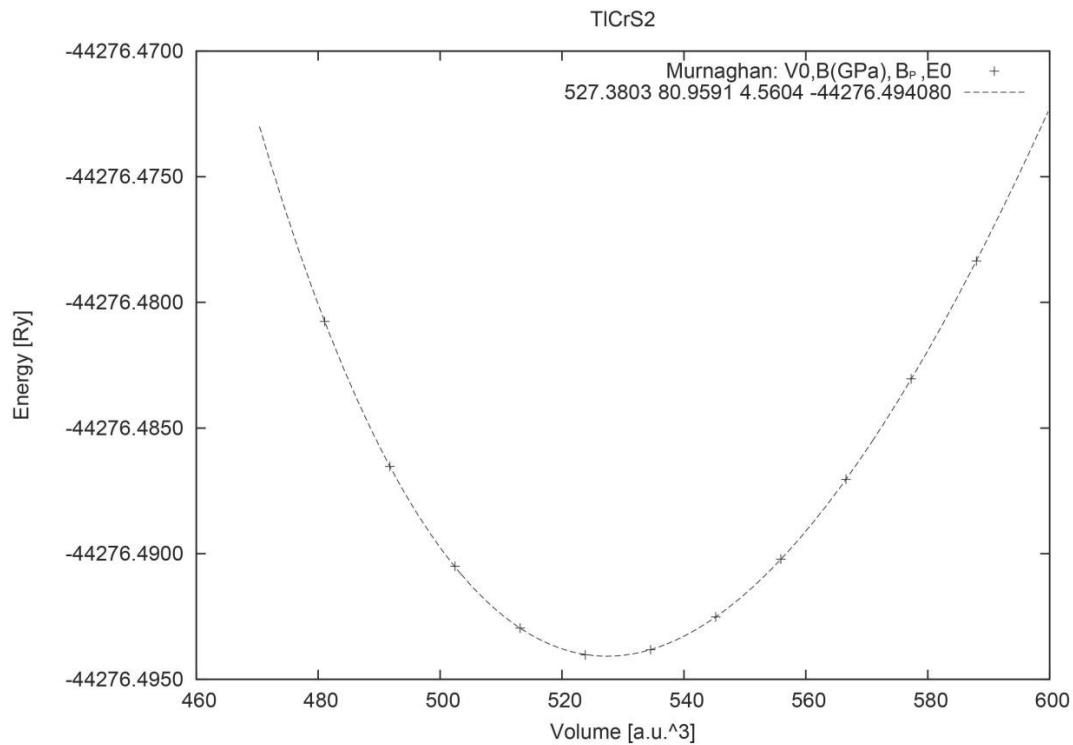

Fig.2. TlCrS$_2$. Volume vs Energy

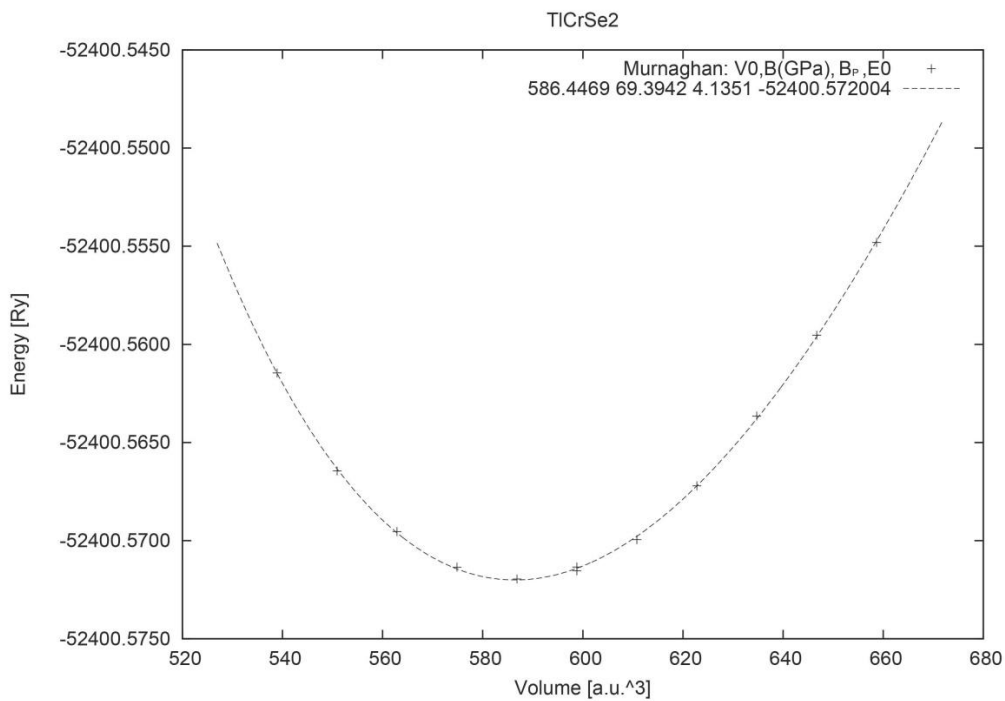

Fig.3. TlCrSe$_2$ . Volume vs Energy

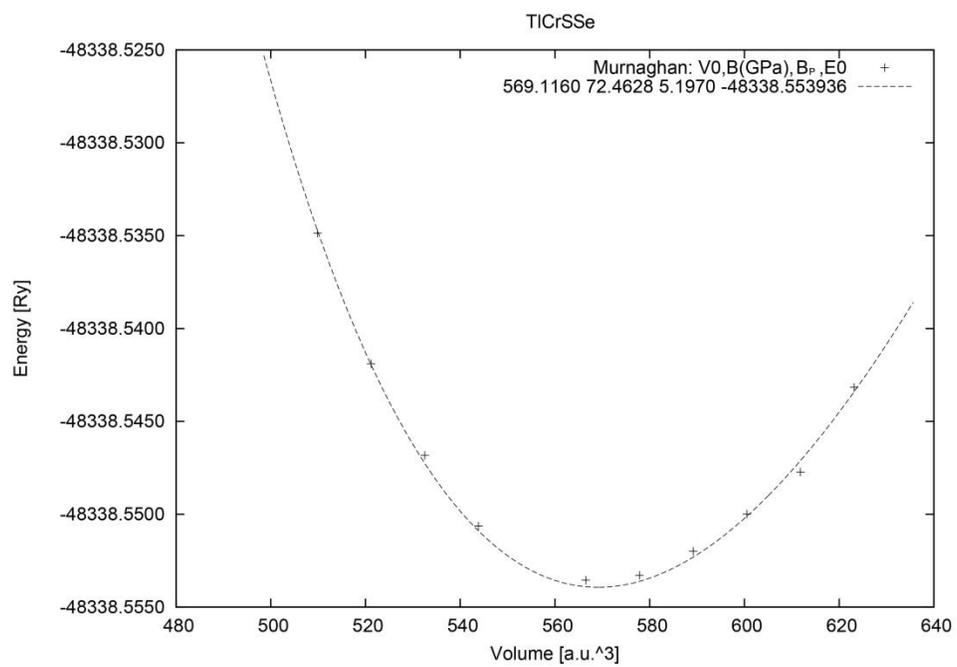

Fig.4. TlCrSSe . Volume vs Energy

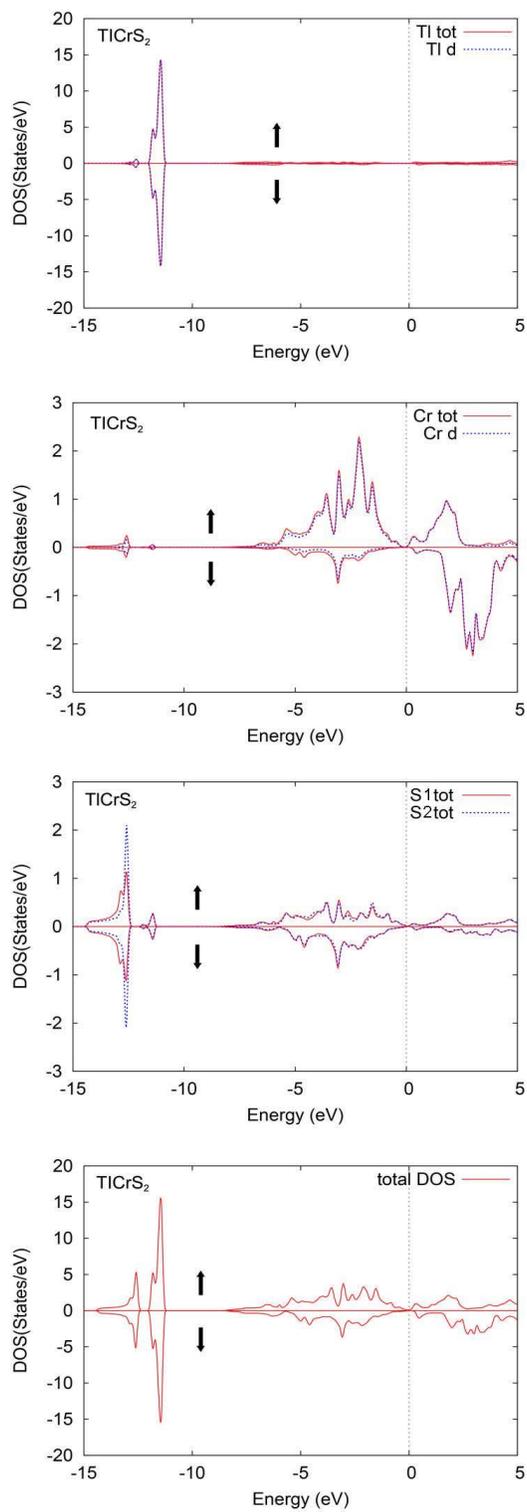

Fig.5. TlCrS$_2$. DOS

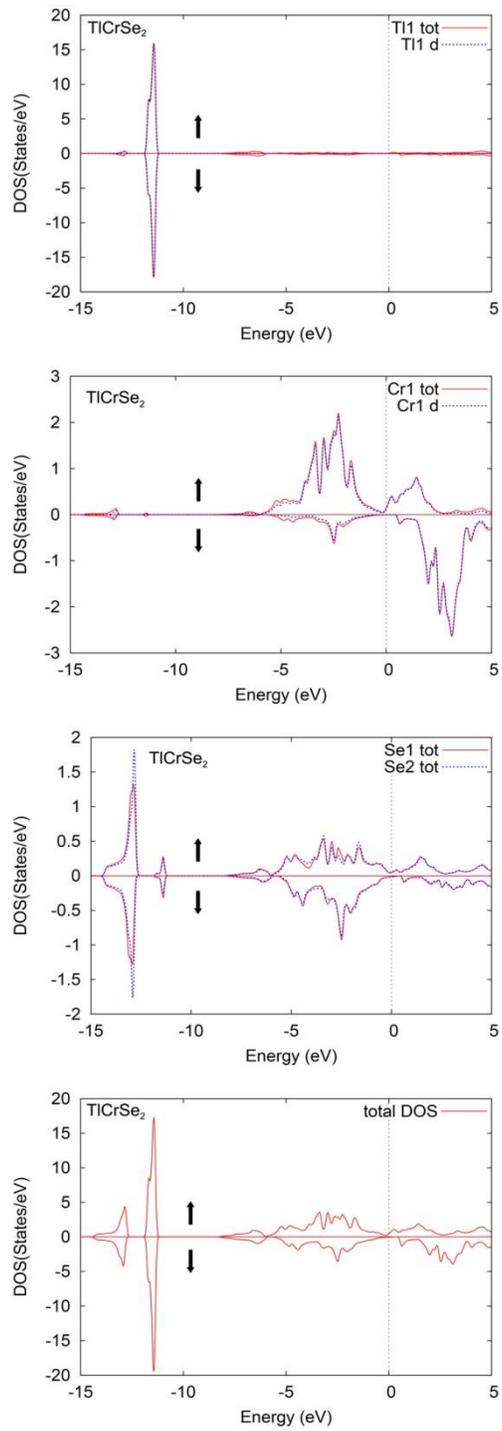

Fig.6.TlCrSe$_2$. DOS

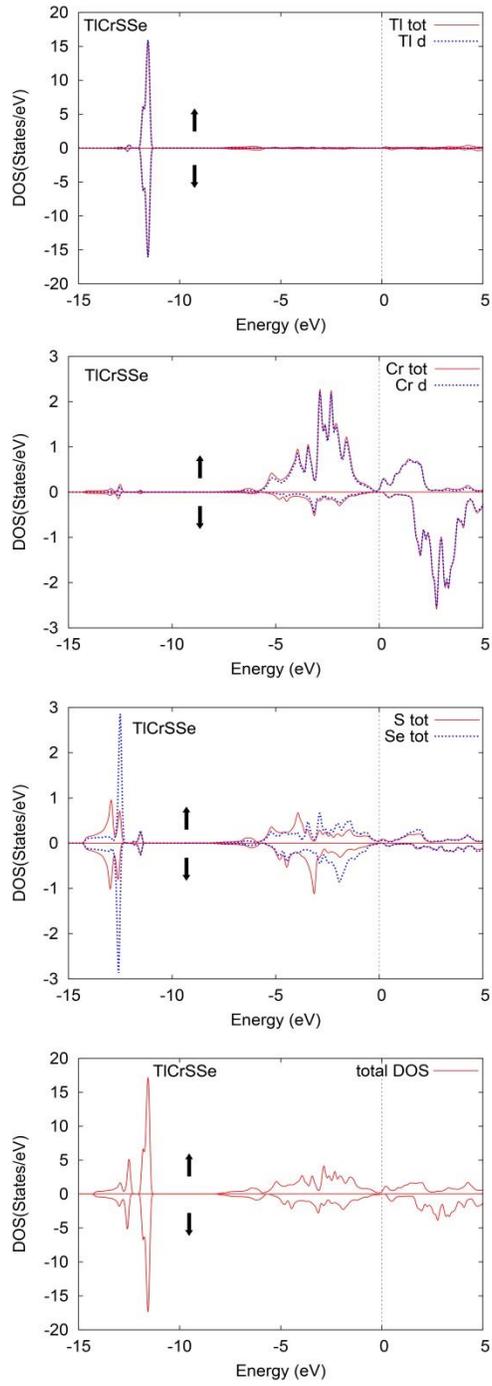

Fig.7.TlCrSSe.DOS

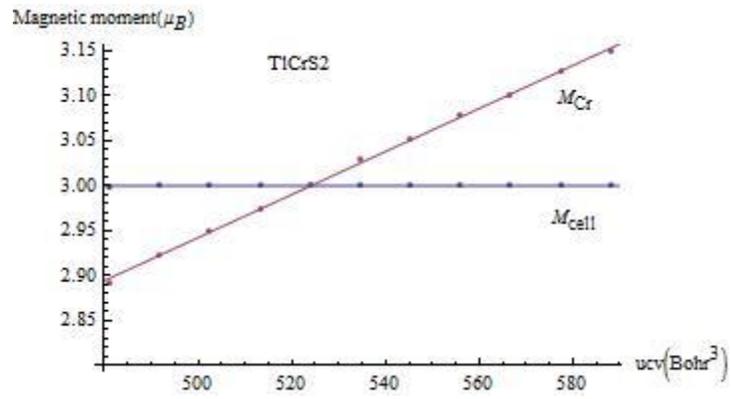

Fig.8.TlCrS$_2$. Volume vs Magnetic moment

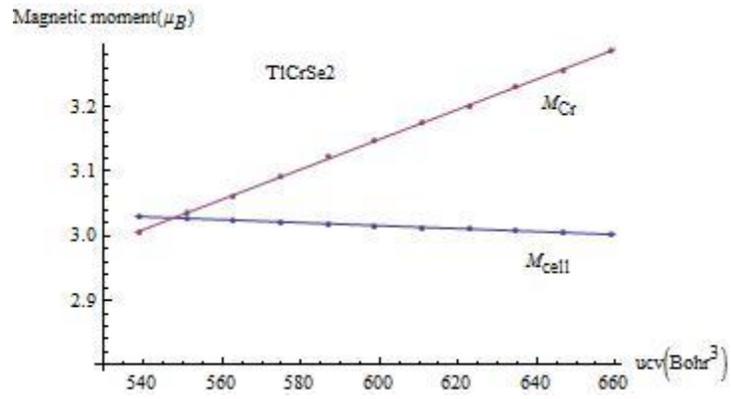

Fig.9.TlCrSe$_2$. Volume vs Magnetic moment

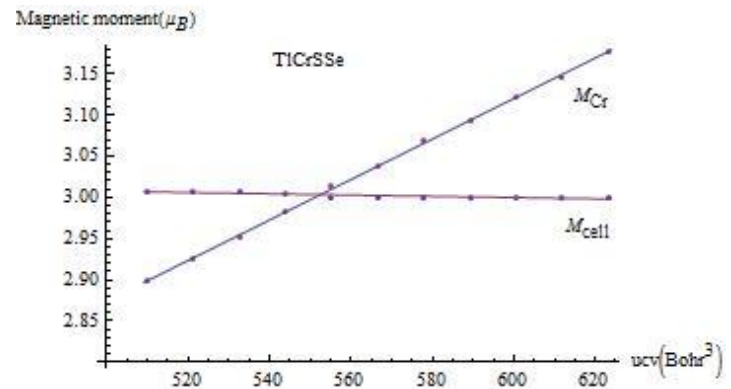

Fig.10.TlCrSSe. Volume vs Magnetic moment